# The impact of social media presence and board member composition on new venture success: Evidences from VC-backed U.S. startups

Gloor, P. A., Fronzetti Colladon, A., Grippa, F., Hadley, B. M., & Woerner, S.

This is the accepted manuscript after the review process, but prior to final layout and copyediting. **Please cite as:**

**The impact of social media presence and board member composition on new venture success: Evidences from VC-backed U.S. startups**

Gloor, P., Fronzetti Colladon, A., Grippa, F., Hadley, B. M., & Woerner, S.

## Abstract

The purpose of this study is to examine the impact of board member composition and board members' social media presence on the performance of startups. Using multiple sources, we compile a unique dataset of about 500 US-based technology startups. We find that startups with more venture capitalists on the board and whose board members are active on Twitter attract additional funding over the years, though they do not generate additional sales. By contrast, startups which have no venture capitalists on the board and whose board members are not on Twitter show an increased ability to translate assets into sales. Consistent with other research, our results indicate that startups potentially benefit from working with VCs because of the opportunity to access additional funding, although their presence does not necessarily translate into sales growth and operational efficiency. We use a number of control variables, including board gender representation and board members' position in the interlocking directorates' network.

## 1. Introduction

Venture capitalists have played a key role in supporting the launch and growth of new companies, especially in the high-tech sector (Cavallo et al., 2019). Venture capital is often considered essential to the growth of new ventures as it provides equity (or equity-linked) investments, facilitates further financial intermediation, and offers managerial direction (Arena et al., 2018; Kortum and Lerner, 2000; Wonglimpiyarat, 2016). Venture capitalists are considered one of the key driving forces in the American entrepreneurial ecosystem (Diana and Ingram, 2014; Phillips, 2018). According to the National Venture Capital Association, venture investors have raised more than $130 billion since 2014 and will continue to deploy capital to high-growth startups (Stanfill et al., 2017).

Several empirical studies have demonstrated the correlation between VC involvement and startup success: VC-backed firms seem to experience faster growth, faster time-to-market of their products, higher productivity, greater innovation, higher efficiency, have more patents, and are more likely to have a successful exit (Bernstain et al., 2016; Chemmanur et al., 2011; Dutta and Folta, 2016; Spender et al., 2017). However, there is little systematic evidence that corporate venture capital investment creates value to investing firms in terms of operational efficiency (Dushnitsky and Lenox, 2006).

This paper extends the work in the area of social legitimacy and signaling (Chaney and Marshall, 2013; Dowling and Pfeffer, 1975). As suggested by Stuart, Hoang and Hybels in their study on endorsements (Stuart et al., 1999), VCs provide legitimacy to the firm as their presence acts as a signal to the public during IPOs. While research has confirmed that board members can be a critical source of advice, providing strategic direction, connections and financial support (Fried et al., 1998; Garg and Eisenhardt, 2017), the relationship between the level of venture capital involvement and startup success has not been systematically scrutinized and there is still room for providing more empirical evidence for the relationship between operational efficiency and venture capital engagement.

While previous studies have focused on the role that VCs hold in supporting new ventures, more research is required to disentangle the specific influence of the VC themselves on the startup and how their social media presence and presence on the board impacts performance. Despite previous studies exploring the association between VCs' presence on the board and positive or negative financial outcomes, findings remain inconclusive. How does the presence of Venture Capitalists on the board of new ventures influence their success? Does it have an impact on specific key performance indicators? In this study, we explore the impact that board of directors and VC presence have in promoting startup growth as we focus in particular on the impact on sales, increase of funding over time and asset turnover. The selection of these indicators is based on the need to account for the new venture's ability to demonstrate efficient use of resources, a sound strategy and high potential for growth by securing additional funding.

Our results provide additional empirical evidence that the specific board composition and the informal social networks created by the members have a direct influence on startups' performance. Among other things, we find that startups which have no venture capitalists on the board and whose board members are not on Twitter show an increased ability to translate assets into sales.

This paper is organized as follows. The first two sections explore the benefits and challenges for startups to actively engage VCs on their board, as well as the impact of VCs' social media presence on the performance of new ventures. These two sections are meant to provide the basis for our theoretical framework, showing that both virtual and physical presence of actively engaged board members and venture capitalists have an impact on new venture success. After introducing and supporting our hypotheses, we focus on describing our sample of 499 startups founded between 2011 and 2016, whose funding data was available on Crunchbase. This analysis is followed by a discussion about the benefits and challenges of inviting venture capitalists to participate in the activities of new ventures.

## 2. Impact of VCs on the Board

Advocates for the positive influence of VCs claim that VCs serve three main roles to identify and promote successful startups: screening, monitoring and coaching (Lahr and Mina, 2016). Screening allows VCs to choose which high quality companies to invest in. VCs are experienced at selecting for certain criteria that predict success, such as technical expertise and founder commitment (Spender et al., 2017). Monitoring requires VCs to track the status of their portfolio companies, comparing investments with market trends and opportunities. They protect the value of their investments by adding credibility and prestige to those companies they invest in (Kaplan and Stromberg, 2003; Reuer and Devarakonda, 2017). Finally, through coaching, VCs provide advice and support to their portfolio companies with the intent of improving their chances of success and, at the end, the return on their investment. This may include connecting the firm with potential resources, assisting with recruitment, and providing experience, advice, and mentoring (Hellmann, 2000; Hellmann and Puri, 2002). For example, a recent study on alliance partner selection (Reuer and Devarakonda, 2017) found that the effects of VC information intermediation are more pronounced when prospective collaborators are at the earliest stages of product development.

Recent studies have empirically demonstrated that inter-organizational networks and social capital represent important measures of the economic performance and growth of a start-up (Pirolo and Presutti, 2010), though it is still debated which social capital configuration is most advantageous for the start-up’s performance during various stages of its life cycle (Borgatti and Cross, 2003; Hite, 2005). Other empirical studies found that human capital and working experience have no significant impact on the success of young high tech firms. As demonstrated by Lasch and colleagues (2007), education level, working experience in small and medium enterprises, start-up training, or launching a business in the same sector as the last employment are not automatically related to success. Factors such as the initial organizational setup of the venture and a long-term financing structure are good predictors of success and sustainable growth. Other studies have highlighted the role of patenting as metric for success of a new venture. For example, Helmers and Rogers (2011) studied high- and medium-tech start-ups in UK and found that the decision to patent positively affects growth in total assets for the between 2001 and 2005.

Kortum and Lerner (2000) examined the impact of venture capital on technological innovation by looking at the patenting patterns across industries over a three-decade period. Their results suggest a positive and significant effect between VC support and industrial innovation. Similarly, Chen (2009) explored the association between venture capital support and technology commercialization and found a moderate improvement effect on the performance: marketing and financial support from VCs seem to matter only when startups possess a lower degree of market scope and a higher degree of technology breadth.

A recent study empirically confirmed the performance benefits of well-matched owners and firms using an extensive longitudinal dataset of the population of U.S. private firms seeking to go public from 1997 to 2004, and their VC owners (Lungeanu and Zajac, 2016). Matching based on VC partners' professional expertise seems to be less dependent on their complementarity with the founder's background and more related to the current lifecycle stage of the startup (Bengtsson and Hsu, 2010).

Other studies have shown that venture capitalists are deeply involved in establishing policies and monitoring managerial activities in high tech firms. The extent of influence is moderated by such factors as stage in the organizational life cycle, size of the venture capitalist, lead position of the venture capitalist, perceived firm performance, background of entrepreneur and relationship between CEO and board members (Garg and Eisenhardt, 2017; Gomez-Mejia et al., 1990).

## 3. Social Media Presence

The value that venture capitalists add to their portfolio companies goes beyond providing consulting, accounting, financial or operational resources. Being associated to the brand of the VC firm provides great value to early stage companies. By definition, startups have no brand at launch, and the connections made available through the VCs' networks is one of their most valuable contributions for an early stage venture (Teten et al., 2013). Gulati and Higgins (2003) found that ties between young biotechnology firms and prominent venture capital firms were valuable to IPO success during cold markets, while ties to prominent investment banks were valuable to IPO success during hot markets. As shown by a study on Taiwanese high-tech new ventures (Lin et al., 2006), successful entrepreneurs are those who can adjust their entrepreneurial strategies according to their social capital and the resources available in their ecosystem.

Organizational scholars have suggested that seed-stage investors rely on social relationships to select which ventures to fund, where their decision is based on two mechanisms: information transfer through social ties and social obligation (Venkataraman, 1997). Shane and Cable (2002) examined the effects on venture finance decisions of direct and indirect ties between entrepreneurs and seed-stage investors. Their field study revealed that entrepreneurs' reputation - defined as information about an individual's past performance - mediates the effects of social ties, demonstrating that investors leverage their social ties to gather private information.

Other studies have stressed the importance of human networks in the formation of business clusters. Myint et al (2005) have highlighted the importance of mini-clusters of key individuals (investors, academics and serial entrepreneurs) to the success of the high technology cluster companies in Cambridge, UK. In particular, they found that startup companies thrive when there is high relational social capital built upon ties of individuals who have worked together in other companies in the past. Furthermore, being physically located in close proximity helps maximize both structural and relational social capital. Here we could argue that social media allows investors and entrepreneurs to establish virtual connections that extend the ability of physical proximity to generate trust and facilitate knowledge sharing (Myint et al., 2005).

Sapienza (1992) demonstrated that the most effective venture capitalists are those who maintain frequent, open communications while minimizing conflicts. The application of a social network approach to study the impact of social capital on new venture success is not new. Hochberg and colleagues (2007) found that when companies are supported by better-networked VCs, they are significantly more likely to survive to subsequent financing and eventual exit.

In a study of the Cambridge high-technology cluster, Myint et al. (2005) used a family tree and interlocking directorships approach to demonstrate the positive impact of social capital on new venture success. An interlocking directorate occurs when an individual affiliated with one organization sits on the board of directors of another organization (Sapinski and Carroll, 2017).

Another recent study explored how entrepreneurs increase social capital in the digital age and looked at how unique technical capabilities of social network sites impact entrepreneurs' bridging and bonding social capital (Smith et al., 2017). Gloor and colleagues (2018) studied the public member profiles of the German business networking site XING to explore the benefits that network position in online business social networks confer to an aspiring entrepreneur. After comparing individual and network attributes in virtual and real-world networks, they found no positive effect of virtual network size and embeddedness on startup success, which seems to confirm that online ties are only as good as the real-world relationships that underlie them.

Only few studies have looked at the impact of social media on new ventures or the role of social media in the capital markets (Akula, 2015). Social media represent an important channel of information for investors to evaluate startup quality through their social media activities. For example, Jin and colleagues (2017) found that startup social media activity is associated with more investment from investors, specifically from those who have access to less information channels such as angels.

Scholars in the area of computer-mediated communications have demonstrated that social capital is accumulated differently online versus offline. While some studies validated the assumption that online friendship networks help entrepreneurs initiate weak ties or manage strong ones (Ellison et al., 2014; Sigfusson and Chetty, 2013), fewer studies have explored how social capital growth in the online networks can help entrepreneurs support their ventures, and what specific impact social capital has on their performance (Gloor et al., 2018). Many have pointed out the importance of human networks and physical proximity (Myint et al., 2005) of entrepreneurs and investors, though more empirical studies are needed to understand how proximity in online social networks affect start-up growth (Batjargal, 2007; Westhead et al., 2003).

## 4. Hypotheses Development

Figure 1 presents the research framework based on the assumption that new ventures are likely to benefit from an active engagement of board members and venture capitalists.

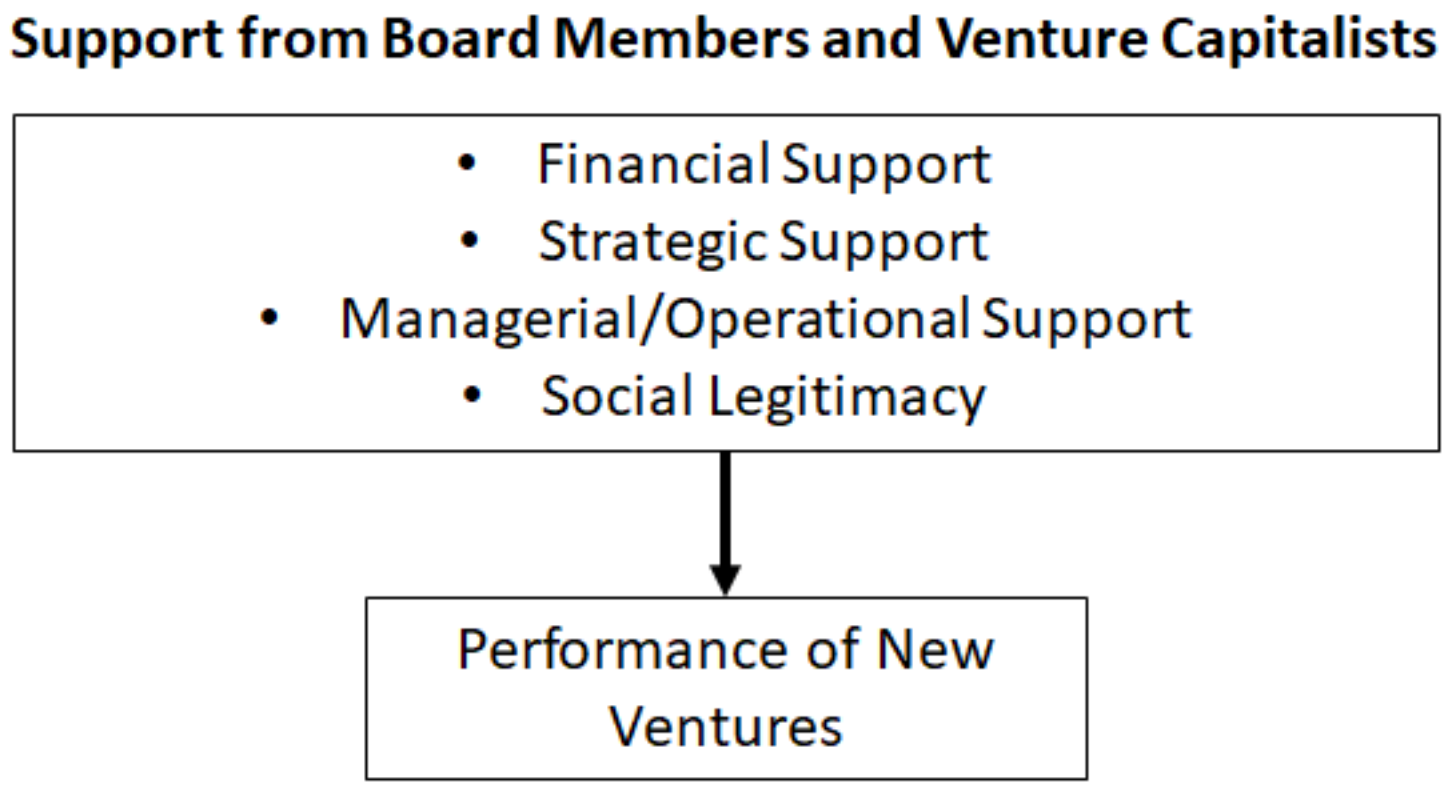

**Figure 1.** Research Framework

Our first hypothesis stems from the assumption that there are more advantages than disadvantages in involving venture capitalists on the board of a startup. As demonstrated by (Stuart et al., 1999) new ventures that are "endorsed" by prominent exchange partners are likely to perform better than otherwise comparable ventures that lack prominent connections. The benefit of having prominent business relationships derives from a transfer of status and legitimacy. The presence of VCs on the board might influence business operations at different levels, as it provides managerial and strategic support that could help entrepreneurs manage resources more efficiently and generate more sales. As demonstrated by previous studies on social legitimacy and the benefits of leveraging inter-organizational ties (Kuhn et al., 2016; Stuart et al., 1999), the presence of VCs on a startup's board helps generate a perception of high quality when other measures of quality are difficult to observe. Based on the signaling theory, we test the first hypothesis by using sales, funding and asset turnover/operational efficiency as proxy for performance.

*H1: A higher number of Venture Capitalists on startups' boards will be positively related to increased performance.*

Besides deciding on the most appropriate VC firm that best matches their new venture, entrepreneurs are often advised to select the right VC partner from that company, based on their understanding of the industry, their operational experience in the startup area, the potential value they can bring and their managerial style and organizational culture. Most importantly, venture capitalists will make a difference based on their ability and willingness to put entrepreneurs in contact with the right partners to help them get into the markets as they build their business.

Startups who can rely on the connections made by their board members on social media are likely to access greater knowledge, improve their understanding of the industry, finding relevant expertise through their board's social ties that can impact their operations and decision-making processes (Smith et al., 2017). Based on the social legitimacy and signaling theory (Chaney and Marshall, 2013; Kibler et al., 2015; Kuhn et al., 2016), we expect that board members' presence on social media will increase the likelihood of increased sales, attracting additional funding. As demonstrated by Kibler et al. (2015) social legitimacy is linked to regional legitimacy among entrepreneurs as this signals a connection to the local community and therefore legitimacy. VCs provide legitimacy to the new ventures, which is further extended by the social legitimacy created when the board is on Twitter. Because the quality of new ventures cannot be easily observed, potential investors are likely to assess the company based on observable attributes that are perceived as associated with its underlying quality (Stuart et al., 1999). Among those "observable attributes" the presence of VCs on social media represents a signal of good quality that would encourage further investment and additional support.

Based on the research conducted by Weill and colleagues (2019), companies with IT savvy boards outperformed others on key metrics including ROA, revenue growth, and market cap growth. Having board members with experience in digital business seems to be the new financial performance differentiator. This translates in our study into the ability to manage social media presence to generate interest and attract managerial and financial resources that are key for new ventures to succeed. This rationale provides support for our second hypothesis in which we state that board members which are more IT savvy and more present on social media, will lead to better performance, operationalized as higher sales, more funding and better asset turnover.

*H2: Presence of board members on Twitter will be positively related to increased performance of the new ventures.*

## 5. Data and Methods

We chose Twitter to study the online communication behavior of venture capitalists sitting on the startup boards, as Twitter is the social media platform most extensively used by startups and investors, and broadly used by the business community (He et al., 2016; Li et al., 2018; Liu et al., 2015). In addition, there are other recent empirical studies that are using Twitter as source of data, which could improve our study's replicability. For example, Cha and colleagues (Cha et al., 2010) analyzed a large dataset on Twitter and explored types and degrees of influence within the network, suggesting that topological measures such as indegree alone (i.e. the number of followers of a user) reveals very little about the influence and popularity of users. Other studies applied both network and sentiment analyses to predict the stock market (Bollen et al., 2011; Zhang et al., 2011).

### 5.1. Sample

The data used for this paper is a sub-set of the sample collected by Hadley et al. (2018), in a study that investigated informal networks of venture capitalists. Hadley collected a dataset extracted from the S&P Capital IQ database, which contained 3199 startups, out of which 1514 had Crunchbase funding data. In our study, we gathered data on a sub-set composed of 499 startups that had OneSource sales data and that were founded between 2011 and 2016. We restricted the sample to startups in the IT/Software domain with headquarters in the United States. We chose to focus on software startups so that we could get access to a large dataset, considering that in 2015 IT new ventures received the largest percentage by sector of all dollars invested in new ventures. This also helped us control for industry effects. The sample included a list of the board members for each startup and specified whether each individual was a venture capitalist or not. Since our focus is on the impact of venture capitalists on startup success, we limited the study to only those firms managing a venture capital investment fund, as opposed to private equity or hedge funds.

### 5.2. Dependent Variables

To assess the startup success we differentiated between the total sales and the total amount of funding that the startup had received since its founding. We collected data on the total amount of funding using Crunchbase (www.crunchbase.com) and data about the startup's sales from OneSource (www.onesource.com). We then calculated the Asset Turnover Ratio (Sales/Total Funding) to assess the efficiency with which a company is deploying its assets in generating revenue. The higher the asset turnover ratio the more efficient the startup is at generating revenue from its asset base. We acknowledge that startup funding, sales, and asset turnover ratio are not absolute measures of startup success, yet in combination they can reflect overall startup success in a meaningful way.

### 5.3. Independent Variables

The first independent variable we included was the engagement of venture capitalists on the board, which has been demonstrated to have a positive impact on firm performance (Eisenhardt and Schoonhoven, 1990; Klotz et al., 2014). The second independent variable is the presence of board members on Twitter.

### 5.4. Control Variables

We included a set of variables that might have an impact on firm performance, which include: gender representation, size and age of the startup, board size, being independent or sponsor-backed, the sub-industry companies belong to, and startups' centrality in the interlocking directorate network.

Several scholars have compared female representation in boardrooms over time (Chen et al., 2016; Farrell and Hersch, 2005) and demonstrated that female representation in top management improves firm performance (Dezsö and Ross, 2012; Post and Byron, 2015). Additional studies suggest that firms operating in industries with greater numbers of female employees are more likely to have female representation on their boards (Bear et al., 2010; Erhardt et al., 2003; Hillman et al., 2007).

We also controlled for startup size and age which have been studied for decades in connection to survival and firm performance (Lotti and Santarelli, 2004; Raz and Gloor, 2007). For example, Baum and colleagues (2000) discussed how and why firm size affects firm performance in a study that indicates the positive influence of establishing alliances on survival, and configuring them into an efficient network.

Another variable that might impact firm profitability is board size. Empirical evidences found negative correlations between board size and profitability (Eisenhardt and Schoonhoven, 1990; Mak and Kusnadi, 2005). Recently, Stam et al. (2014) found that the impact of entrepreneurs' personal networks on performance was mediated by the age of small firms, the industry and institutional contexts in which the small firms operate, and by the specific network or performance measures used. Since the size of the board could also impact performance, we also included board size as another control variable. Boards with a large number of members might experience low efficiency due to increased coordination mechanism costs, where small size boards could be more agile and reap better opportunities faster than slow-moving boards. At the same time, startups with fewer board members might be deprived of important connections, resources and managerial support that large boards provide to startups.

We also controlled for the level of support received by startups, based on empirical evidences that group firms or firms that are backed by others may have higher growth as they can draw on technology and organizational expertise from other firms (Zahra, 1996). Baum et al. (2000) found that independent ventures or corporate sponsored ventures are influenced by technology strategies that may affect their development and success.

Given the widely recognized benefits of social networks in providing access to critical resources such as financial or legal services, or to business partnerships, we controlled for the network position of startups based on the number of board members they share. To control for the impact of real-world ties, we built a formal network of interlocking directorates based on companies (the nodes) who were linked to other companies due to the same director sitting on the boards of both companies. Interlocking directorates represent a good proxy for real-world ties, even if indirect ones. The benefits of being connected with other entrepreneurs and investors, either through strong or weak ties, has been widely recognized in literature (Battilana and Casciaro, 2012). Specifically, we analyzed the social position of startups in the interlocking network by means of standard centrality algorithms, calculating the measures of betweenness and degree centrality (Wasserman and Faust, 1994). In particular, degree centrality counts the number of direct connections a node has in the social network, with startups that share more board members having higher scores of this metric. Betweenness centrality, on the other hand, measures how many times a social actor is in-between the shortest network paths that interconnect the other nodes. Startups with higher betweenness centrality exhibit connections that span across different social groups – they often act as mediators, have more brokerage power and easier access to business resources (Allen et al., 2016).

Finally, we controlled for the industry group within the Information Technology sector (Kile and Phillips, 2009), by differentiating among the following categories: Semiconductors, Internet Software and Services, Data Processing and Outsourced Services, Home Entertainment Software, Application Software, Electronic Equipment and Instruments, Communications Equipment, IT Consulting and Other Services, Electronic Components, Semiconductor Equipment, Technology Hardware, Storage and Peripherals.

Figure 2 illustrates the variables used in the study to explain the connections between number of board members and performance as well as members' social media activity and performance.

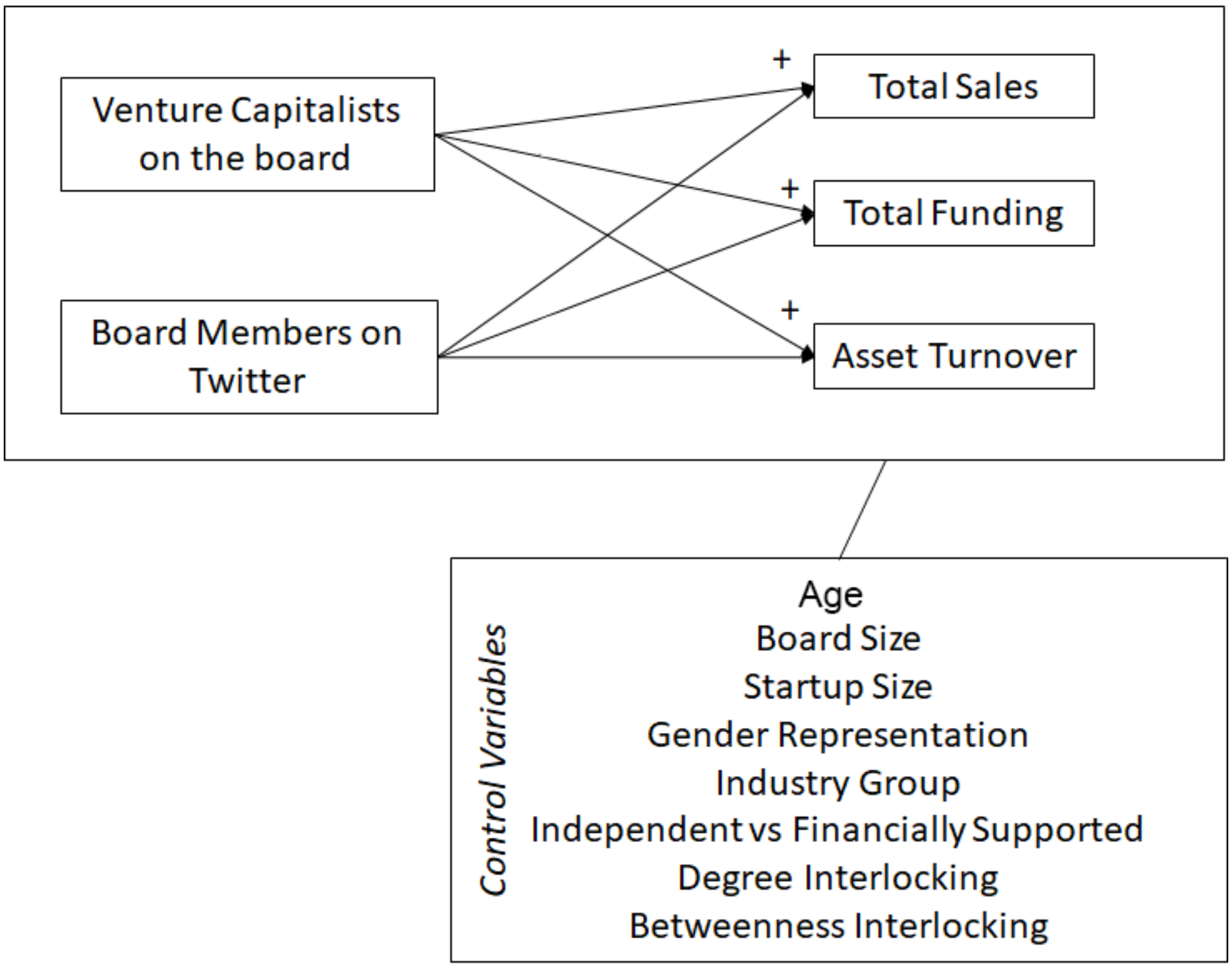

**Figure 2.** Variables and Hypotheses

## 6. Results

Table 1 presents the descriptive statistics for our variables. It is worth noting that the average percentage of female board members is rather low (7.4%). The maximum number of board members is 10 (3.45 on average), and the average degree of interlocking is 4.86. Only 12% of startups was independent, all the others being sponsor-backed. 57.1% of the startups had at least one VC among their board members and 57.9% had at least one board member on Twitter. Startups' size ranged from 1 to 602 employees, while startups' age was anywhere between less than a year and 5 years.

| Variable | M | SD | Min | Max |
|---|---|---|---|---|
| Sales (USD Millions) | 10.36 | 19.23 | .1 | 245.10 |
| Funding (USD Millions) | 26.84 | 102.99 | .03 | 1779 |
| Number of VC on the Board | .91 | 1.01 | 0 | 5 |
| Board Members on Twitter | .58 | .49 | 0 | 1 |
| Startup Size | 30.95 | 52.60 | 1 | 602 |
| Startup Age | 4.07 | 1.01 | 0 | 5 |
| Board Size | 3.45 | 2.38 | 1 | 10 |
| Independent (no previous financial support) | .12 | .32 | 0 | 1 |
| Females Percentage in the Board | .07 | .23 | 0 | 1 |
| Betweenness Interlocking | 8497.45 | 11029.61 | 0 | 79789.48 |
| Degree Interlocking | 4.86 | 3.48 | 0 | 22 |

**Table 1**. Descriptive Statistics

In Table 2 and Table 3, we present the multiple regression models used to assess the impact of our independent variables on sales and funding (dependent variables were logarithmically transformed).

| Variable | Model 1 | Model 2 | Model 3 | Model 4 | Model 5 | Model 6 | Model 7 |
|---|---|---|---|---|---|---|---|
| **Number of VC on the Board** | | .60288*** | | | | .30945*** | .29856*** |
| **Board Members on Twitter** | | | | 1.24405*** | | .59208*** | .57468*** |
| **Startup Size** | .01060*** | | | | | .00918*** | .00878*** |
| **Startup Age** | -.029972 | | | | | | -.04128 |
| **Board Size** | .05716* | | | | | | -.12628* |
| **Independent (no previous financial support)** | -1.70964*** | | | | | -1.22601*** | -1.18816*** |
| **Females Percentage in the Board** | | .13777 | | | | | .12974 |
| **Betweenness Interlocking** | | | -.00003** | | | | .00001 |
| **Degree Interlocking** | | | .22967*** | | | | .05364 |
| **Semiconductors** | | | | | -.37029 | .05204 | .16381 |
| **Internet Software and Services** | | | | | .35076 | .27385 | .31330 |
| **Data Processing and Outsourced Services** | | | | | .55276 | .66091 | .71248 |
| **Home Entertainment Software** | | | | | -.26864 | .16102 | .22375 |
| **Application Software** | | | | | .44422 | .40944 | .47309 |
| **Electronic Equipment and Instruments** | | | | | .13232 | .28944 | .38713 |
| **Communications Equipment** | | | | | .06837 | .62039 | .69646 |
| **IT Consulting and Other Services** | | | | | -2.28967* | -.96259 | -.93384 |
| **Electronic Components** | | | | | .66778 | .04184 | .04676 |
| **Semiconductor Equipment** | | | | | -1.39570 | -.68247 | -.57874 |
| **Technology Hardware, Storage and Peripherals** | | | | | 1.50011* | 1.14420 2* | 1.15267* |
| **Constant** | 2.06478*** | 1.70441*** | 1.43969*** | 1.54508*** | 1.95531*** | 1.20535*** | 1.43836*** |
| **Adjusted R-squared** | .3163 | .1653 | .1173 | .1719 | .0228 | .4199 | .4260 |
| **N** | 499 | 499 | 499 | 499 | 499 | 499 | 499 |

*p < .05; **p < .01; ***p < .001

**Table 2**. Predictors of a startup's ability to attract funding (USD millions).

| Variable | Model 1 | Model 2 | Model 3 | Model 4 | Model 5 | Model 6 | Model 7 |
|---|---|---|---|---|---|---|---|
| **Number of VC on the Board** | | .14558*** | | | | | .03240 |
| **Board Members on Twitter** | | | | .31721*** | | .13037** | .12294** |
| **Startup Size** | .00715*** | | | | | .00685*** | .00677*** |
| **Startup Age** | -.00494 | | | | | | -.00036 |
| **Board Size** | .01693* | | | | | | -.03487 |
| **Independent (no previous financial support)** | -.29509*** | | | | | -.22201** | -.20378** |
| **Females Percentage in the Board** | | .13486 | | | | | .14331 |
| **Betweenness Interlocking** | | | -.00001 | | | | .00001 |
| **Degree Interlocking** | | | .06947*** | | | .01513* | .01551 |
| **Semiconductors** | | | | | .09679 | | .26349 |
| **Internet Software and Services** | | | | | .06298 | | .01682 |
| **Data Processing and Outsourced Services** | | | | | .26511 | | .28591 |
| **Home Entertainment Software** | | | | | -.24863 | | -.12168 |
| **Application Software** | | | | | .10120 | | .07813 |
| **Electronic Equipment and Instruments** | | | | | -.02218 | | .07084 |
| **Communications Equipment** | | | | | -.15674 | | .03697 |
| **IT Consulting and Other Services** | | | | | -.47919 | | -.11157 |
| **Electronic Components** | | | | | .32504 | | .00469 |
| **Semiconductor Equipment** | | | | | .53580 | | -.25647 |
| **Technology Hardware, Storage, Peripherals** | | | | | .40679 | | .25620 |
| **Constant** | .41194*** | .49343*** | .37479*** | .45278*** | .57720*** | .30208*** | .30260* |
| **Adjusted R-squared** | .4440 | .0562 | .0732 | .0671 | .0038 | .4635 | .4679 |
| **N** | 499 | 499 | 499 | 499 | 499 | 499 | 499 |

*p < .05**;** **p < .01; *****p < .001**

**Table 3**. Predictors of a startup's ability to increase sales (USD millions).

Predictors were first tested in blocks and then combined in Model 6, if significant together. Model 7 is the full model in both tables. We refer to Model 6 as the best model, as it is more parsimonious. Moreover, in Model 7 degree centrality in the interlocking network is partially collinear with the number of board members (maximum VIF is 7.39). Our models show that depending on which performance indicator we consider, board member composition and board members' presence on Twitter have a different impact on startup success. Startups with more venture capitalists on the board tend to attract additional funding. This, however, does not translate into greater sales. Startups experience more funding and increased sales when their board members are on Twitter (even if the effect on sales is smaller). In addition, a higher degree centrality of board members in the interlocking network seems to be beneficial to sales, but this predictor is not significant in the full model. When controlling for the industry group our results experience little change. We only find that more funding is attracted by startups producing technology hardware, peripherals and storage solutions.

Another interesting result is that having a board with several venture capitalists and having many of them active on Twitter does not guarantee an increased ability to manage their resources efficiently. This result is presented in the ANOVA table and figure (Table 4 and Figure 3): startups with a higher operational efficiency are the ones who have no venture capitalists on the board and whose board members are not on Twitter.

| | | | | | | | **Post hoc analysis (Tukey HSD)** | | | |
|---|---|---|---|---|---|---|---|---|---|---|
| | **Sum of Squares** | **df** | **Mean Square** | **F** | **Sig.** | **Mean Asset Turnover** | **G1** | **G2** | **G3** | **G4** |
| **Between groups** | 364.542 | 3 | 121.514 | 5.253 | .001 | G1 = .552 | | | | *** |
| **Within groups** | 11451.023 | 495 | 23.133 | | | G2 = 1.334 | | | | |
| **Total** | 11815.566 | 498 | | | | G3 = 1.232 | | | | |
| | | | | | | G4 = 2.635 | *** | | | |

*Note.* ***p < .001. G1 = Firms with VCs in the board and with board members on Twitter; G2 = Firms with VCs in the board and without board members on Twitter; G3 = Firms without VCs on the board and with board members on Twitter; G4 = Firms without VCs on the board and without board members on Twitter.

**Table 4. One-way** ANOVA of Asset Turnover and post-hoc tests (Tukey HSD)

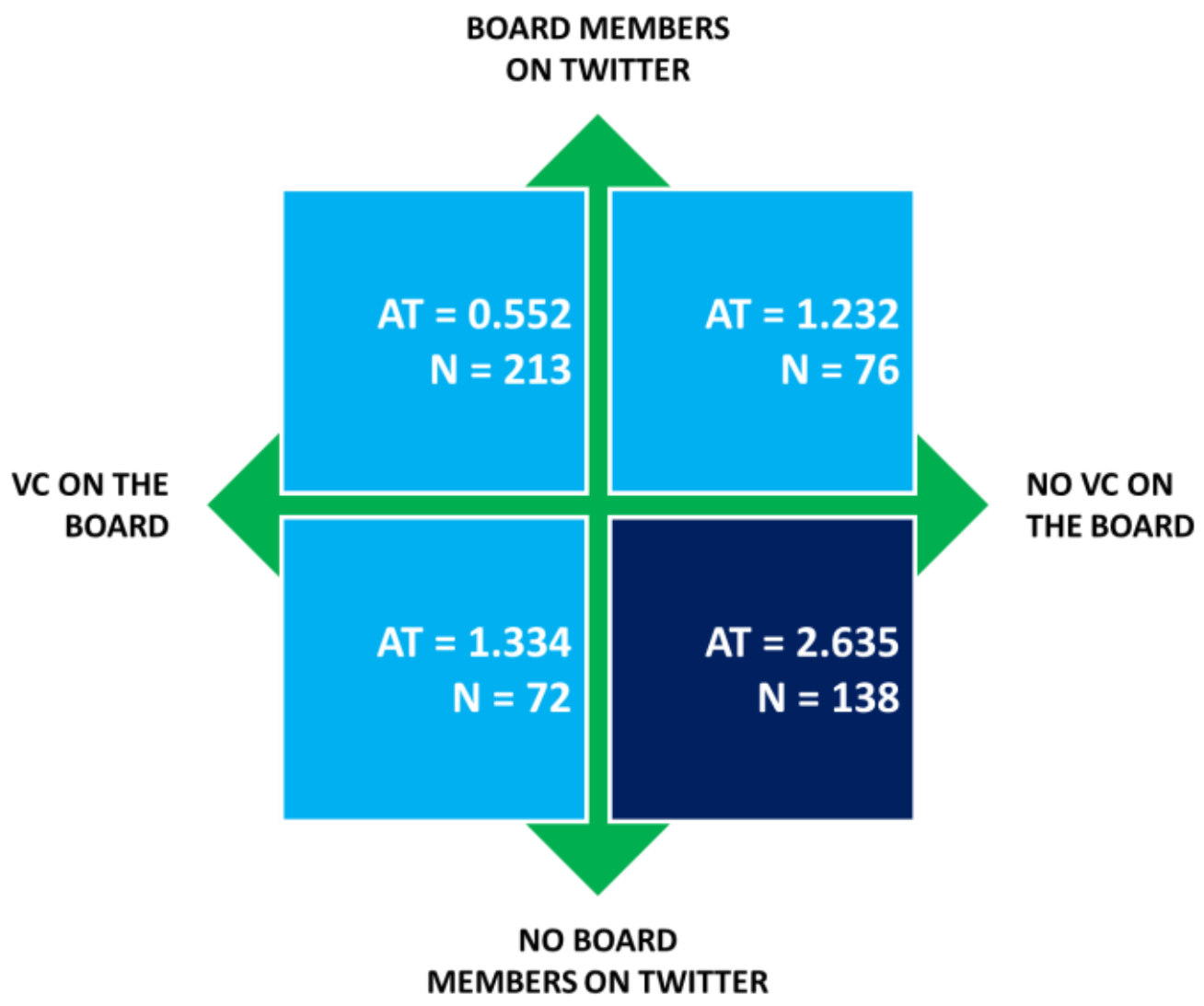

**Figure 3.** Results of the ANOVA explaining the impact on Asset Turnover (AT) of board members on Twitter and VCs on the board

Figure 4 summarizes the factors impacting startup success based on our models. The ability to attract additional funding seems to be associated with having VCs on the board. Similarly, our second hypothesis has been only partially confirmed since a higher number of Venture Capitalists on startups' boards are not positively related to increased asset turnover. As suggested by the social signaling theory, social media presence might impact the ability to attract additional funding and help define a winning sales strategy (Chaney and Marshall, 2013; Stuart et al., 1999) though it might not be enough to improve the operational efficiency of the new ventures.

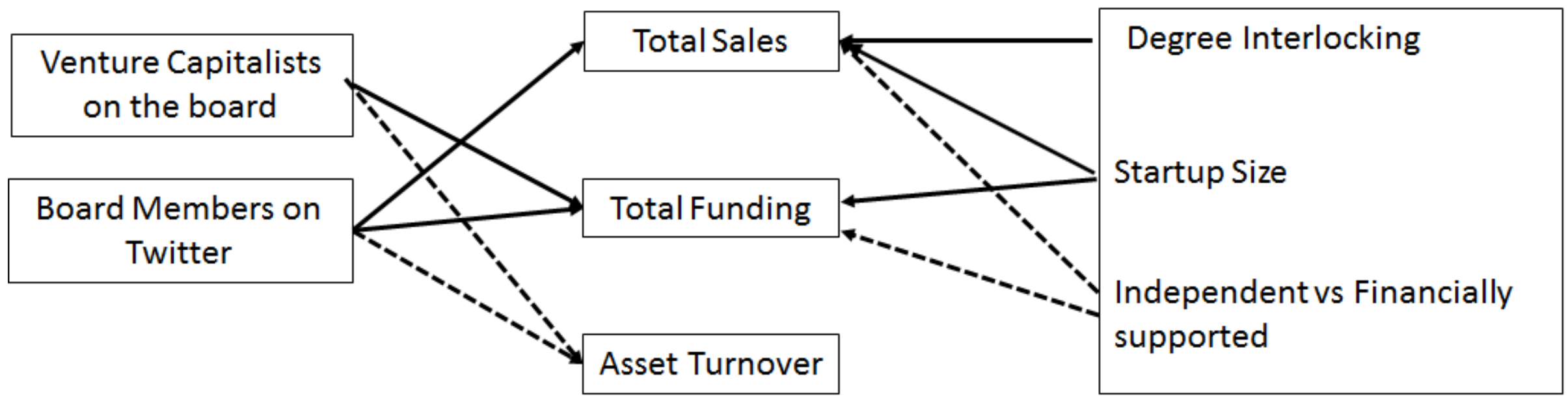

**Figure 4**. Factors impacting startup performance. Solid arrows indicate positive influence. Dotted arrows indicate negative influence.

## 7. Discussion

The results indicate that startup performance can be explained only partially by VCs participation on the board and by the activity on social media of the board members. Our models partially confirm the benefit of inviting venture capitalists to participate in the activities of new ventures. Inviting VCs with the right connections, including the ones with strong ties on social media, seems to have a positive effect on funding, as they benefit from the VCs' brand equity and may access additional funding by virtue of being linked to the right investors. This is aligned with previous studies on the benefits of inter-organizational connections and social signaling, which show that connections made available through the VCs' networks is one of their most valuable contributions for an early stage venture (Teten et al., 2013).

Given the evidence suggesting that venture capitalists support startup growth by offering managerial and financial support, we expected an increased ability to manage resources efficiently. Contrary to this assumption, our hypothesis (H1) was not confirmed as shown in the ANOVA. This means that neither VCs were able to positively assist startups making efficient use of resources, nor did their presence significantly help increase sales. We speculate that startup founders were ill advised on how to best utilize their assets or what investment decisions would be more profitable to grow their company. Another possible reason for not being able to deploy their assets in generating more revenue could be the emergence of conflicts within the board. While there seems to be a widespread theoretical assumption that a business is more successful when several individuals are synergistically combining their skills and resources, both tangible and intangible (Eisenhardt and Schoonhoven, 1990; Klotz et al., 2014), Greenberg and Mollick (2018) found that ventures started by solo entrepreneurs generally outperform teams of co-founders, particularly two-person teams.

We also found that startup size is predictive of increased performance for both indicators (sales and funding). This can be explained by considering that startups with a larger number of employees have better recognition on the market, easier access to additional investors, can rely on additional sources of knowledge through talent acquisition as well as complementary assets such as a network of suppliers that can boost their sales (Raz and Gloor, 2007).

In addition, we found that the degree of interlocking might be predictive of the company's ability to generate more sales: startups that share at least one board member have a better opportunity to grow their sale opportunities. This effect is aligned with previous research, for example with a study of 90 new ventures in the open source software industry indicating that the combination of high network centrality and extensive bridging ties strengthened the link between entrepreneurial orientation and performance (Stam and Elfring, 2008).

Lastly, our results show that startups whose board members are active on Twitter have increased sales and funding. However, this social media presence does not seem to conduct to increased performance, measured as asset turnover. This is consistent with past research. For example, in a study of one million public member profiles of the German business networking site Xing, Gloor and colleagues (2018) found that business performance was not affected by virtual network size and embeddedness.

## 8. Managerial and Theoretical Implications

While it has been clearly demonstrated that VCs on the boards offer financial, intellectual, and social capital resources (Reuer and Devarakonda, 2017; Spender et al., 2017) our study provides additional empirical evidence that the specific board composition and the informal social networks created by the members have a direct influence on startups' performance.

Since board members often sit on multiple boards, they may share their resources with multiple startups, and facilitate a knowledge sharing process that could yield benefits for founders and investors.

Recent evidences suggest that venture capitalists might have a negative influence on startups and their entrepreneurs, especially when conflicts arise between startup entrepreneurs and VCs (Higashide and Birley, 2002). The VC business model has been recently criticized by entrepreneurs who feel that, by encouraging new ventures to expand too quickly, venture capitalists can make them “accelerate straight into the ground” (Griffith, 2019).

Prior research has identified three main areas of VC-CEO conflicts: conflicts of interests and unfavorable attributions, conflicts of inefficient collaboration, and conflicts of VC-CEO mismatch (Khanin and Turel, 2013). Conflict does not necessarily have a negative impact on the success of a startup (Higashide and Birley, 2002), though at high levels of occurrence and intensity, conflict is generally considered to be costly to those involved. Recently growing skepticism has been voiced in the Silicon Valley ecosystem encouraging entrepreneurs to be cautious of venture capitalist investment (Mullin, 2014). To mitigate the tradeoff of resource available versus retained power, entrepreneurs can take funding only when they are able to protect their intellectual property or when able to retain control over their operations.

While neither board size, nor the presence of more female members, impact the ability to increase sales or generate external funding, interesting results emerge if we consider the effect of previous financial support from other investors. New ventures that received previous financial backing experienced greater sales and attracted additional funding. This could be explained by the positive effect that brand awareness of the VC companies has on their portfolio of companies: the sooner a startup is able to access a network of investors, the higher its chances to benefit from the brand awareness of the VCs and convince additional investors that it is worth investing in.

While there is no doubt that investors' social networks bring positive effects to the capability to draw additional funding, it is not always clear whether the inclusion of the investors' perspective positively influences the ability to properly allocate resources and make wise investment decisions. Venture capitalists certainly offer important financial expertise and provide access to invaluable financing opportunities that startups desperately need in their early stages of growth. At the same time, entrepreneurs might be better off making decisions without the often self-centered influence of VCs who bring a partial perspective biased by their own self-interest. As our results on the degree of interlocking demonstrate, the more startups share VCs on their board, the more sales they are likely to generate. This comes at a cost, though, since VCs might have undisclosed conflict of interests and might not be able to offer an unbiased picture of the industry trends. Again, it is a trade-off between the benefits of going solo, including a desire to keep control of the company, and the costs associated with lack of connections to industry networks and the risk of growing at a slower pace (Greenberg and Mollick, 2018). Relinquishing control is one of the main reasons many high-tech companies have refused to go public in the past and successful entrepreneurs are the ones who are able to balance between engaging with venture capitalists and their networks and demonstrating business acumen, without being overly influenced by the investors. As new ventures move from one stage of growth to the next, entrepreneurs face key problems such as how to finance the growth and how to identify the right assets to use. As they search for managers to help the company grow, they can tap into the social networks of venture capitalists to hire managers with an eye to the company's future. Being supported by VCs is not a guarantee of success, though. Entrepreneurs are advised to keep on investing in the right talent, delegate responsibilities to others to improve the managerial effectiveness of a fast growing company. While leveraging investors' connections, online and in person, could help boost sales and attract additional funding thanks to the marketing network externalities involved, entrepreneurs should keep control of the operational mechanisms that generate value by transforming resources into products in the most efficient and effective way.

Lastly, by observing the role and behavior of VCs and board members on Twitter, our study provides a unique contribution to the literature, as it offers a new approach to combine formal and informal social networks to study their impact on startup success. By focusing on the online and real-world ties of board members, we offer additional insights to the conversation on the role of online communication and weak ties in shaping the success of startups. While previous studies found that online communication may be associated with business growth (Kuhn et al., 2016), our study offers a different lens on the type of impact that these ties can bring. While real-world ties, in the form of previous investment support or interlocking directorates, might have an impact on sales and additional funding, these ties are not necessarily translated into the entrepreneurs' ability to create value out of the assets available to them.

## 9. Conclusions and Limitations

While the empirical results shed an interesting light on the influence of social media presence on startup success, caution should be exerted when generalizing the findings. The limitations mainly originate from the choice of analyzing one single industry and to test online presence considering a single social media platform (Twitter). Further research could extend our findings, considering a larger sample of startups, operating in different business sectors and/or in countries other than the US. Online activity could also be tested on other media, such as Facebook, LinkedIn or specialized business forums and blogs. Previous studies have not provided consistent conclusions regarding the impact of VC investments on entrepreneurial firms, and whether this impact is positive or negative. Our study offers the opportunity to reflect on the actual impact that investors have when they join the board of directors in newly created companies.

## References

Akula, S.C., 2015. The Influence of Social Media Platforms for Startups. Journal of Mass Communication & Journalism 5, 1–4. https://doi.org/10.4172/2165-7912.1000264

Allen, T.J., Gloor, P., Fronzetti Colladon, A., Woerner, S.L., Raz, O., 2016. The power of reciprocal knowledge sharing relationships for startup success. Journal of Small Business and Enterprise Development 23, 636–651. https://doi.org/10.1108/JSBED-08-2015-0110

Arena, M., Bengo, I., Calderini, M., Chiodo, V., 2018. Unlocking finance for social tech start-ups: Is there a new opportunity space? Technological Forecasting and Social Change 127, 154–165. https://doi.org/10.1016/j.techfore.2017.05.035

Batjargal, B., 2007. Internet entrepreneurship: Social capital, human capital, and performance of Internet ventures in China. Research Policy 36, 605–618. https://doi.org/10.1016/j.respol.2006.09.029

Battilana, J., Casciaro, T., 2012. Change agents, networks, and institutions: A contingency theory of organizational change. Academy of Management Journal. https://doi.org/10.5465/amj.2009.0891

Baum, J.A.C., Calabrese, T., Silverman, B.S., 2000. Dont go it alone: Alliance network composition and startups’ performance in Canadian biotechnology. Strategic Management Journal 21, 267–294. https://doi.org/10.1002/(SICI)1097-0266(200003)21:3<267::AID-SMJ89>3.0.CO;2-8

Bear, S., Rahman, N., Post, C., 2010. The Impact of Board Diversity and Gender Composition on Corporate Social Responsibility and Firm Reputation. Journal of Business Ethics 97,

207–221. https://doi.org/10.1007/s10551-010-0505-2

Bengtsson, O., Hsu, D.H., 2010. How Do Venture Capital Partners Match with Startup Founders? SSRN Electronic Journal. https://doi.org/10.2139/ssrn.1568131

Bernstain, S., Giroud, X., Townsend, R.R., 2016. The Impact of Venture Capital Monitoring. The Journal of Finance 71, 1591–1622. https://doi.org/10.1111/jofi.12370

Bollen, J., Mao, H., Zeng, X., 2011. Twitter mood predicts the stock market. Journal of Computational Science. https://doi.org/10.1016/j.jocs.2010.12.007

Borgatti, S.P., Cross, R., 2003. A Relational View of Information Seeking and Learning in Social Networks. Management Science 49, 432–445. https://doi.org/10.1287/mnsc.49.4.432.14428

Cavallo, A., Ghezzi, A., Dell'Era, C., Pellizzoni, E., 2019. Fostering digital entrepreneurship from startup to scaleup: The role of venture capital funds and angel groups. Technological Forecasting and Social Change 145, 24–35. https://doi.org/10.1016/j.techfore.2019.04.022

Cha, M., Haddai, H., Benevenuto, F., Gummadi, K.P., 2010. Measuring User Influence in Twitter : The Million Follower Fallacy. International AAAI Conference on Weblogs and Social Media 10–17. https://doi.org/10.1.1.167.192

Chaney, D., Marshall, R., 2013. Social legitimacy versus distinctiveness: Mapping the place of consumers in the mental representations of managers in an institutionalized environment. Journal of Business Research. https://doi.org/10.1016/j.jbusres.2012.09.018

Chemmanur, T.J., Krishnan, K., Nandy, D.K., 2011. How Does Venture Capital Financing Improve Efficiency in Private Firms? A Look Beneath the Surface. Review of Financial Studies 24, 4037–4090. https://doi.org/10.1093/rfs/hhr096

Chen, C.J., 2009. Technology commercialization, incubator and venture capital, and new venture performance. Journal of Business Research 62, 93–103.

https://doi.org/10.1016/j.jbusres.2008.01.003

Chen, G., Crossland, C., Huang, S., 2016. Female board representation and corporate acquisition intensity. Strategic Management Journal. https://doi.org/10.1002/smj.2323

Dezsö, C.L., Ross, D.G., 2012. Does female representation in top management improve firm performance? A panel data investigation. Strategic Management Journal 33, 1072–1089. https://doi.org/10.1002/smj.1955

Diana, M.H., Ingram, A., 2014. A review of the entreprenurial ecosystem and the entrepreneurial society in the United States: An exploration with the global entrepreneurship monitor dataset. Journal of Business and Entrepreneurship.

Dowling, J., Pfeffer, J., 1975. Organizational Legitimacy: Social Values and Organizational Behavior. The Pacific Sociological Review. https://doi.org/10.1039/F29817702213

Dushnitsky, G., Lenox, M.J., 2006. When does corporate venture capital investment create firm value? Journal of Business Venturing 21, 753–772. https://doi.org/10.1016/j.jbusvent.2005.04.012

Dutta, S., Folta, T.B., 2016. A comparison of the effect of angels and venture capitalists on innovation and value creation. Journal of Business Venturing 31, 39–54. https://doi.org/10.1016/j.jbusvent.2015.08.003

Eisenhardt, K.M., Schoonhoven, C.B., 1990. Growth : Organizational Linking Founding Team , Strategy , Environment , and Growth among U . S . Semiconductor. Administrative Science Quarterly 35, 504–529. https://doi.org/10.2307/2393315

Ellison, N.B., Vitak, J., Gray, R., Lampe, C., 2014. Cultivating social resources on social network sites: Facebook relationship maintenance behaviors and their role in social capital processes. Journal of Computer-Mediated Communication. https://doi.org/10.1111/jcc4.12078

Erhardt, N.L., Werbel, J.D., Shrader, C.B., 2003. Board of Director Diversity and Firm

Financial Performance. Corporate Governance 11, 102–111. https://doi.org/10.1111/1467-8683.00011

Farrell, K.A., Hersch, P.L., 2005. Additions to corporate boards: The effect of gender. Journal of Corporate Finance. https://doi.org/10.1016/j.jcorpfin.2003.12.001

Fried, V.H., Bruton, G.D., Hisrich, R.D., 1998. Strategy and the board of directors in venture capital-backed firms. Journal of Business Venturing 13, 493–503. https://doi.org/10.1016/S0883-9026(97)00062-1

Garg, S., Eisenhardt, K.M., 2017. Unpacking the CEO-Board relationship: How strategy making happens in entrepreneurial firms. Academy of Management Journal 60, 1828–1858. https://doi.org/10.5465/amj.2014.0599

Gloor, P.A., Woerner, S., Schoder, D., Fischbach, K., Colladon, A.F., 2018. Size does not matter - in the virtual world. Comparing online social networking behaviour with business success of entrepreneurs. International Journal of Entrepreneurial Venturing 10, 435–455. https://doi.org/10.1504/IJEV.2018.093919

Gomez-Mejia, L.R., Balkin, D.B., Welbourne, T.M., 1990. Influence of venture capitalists on high tech management. Journal of High Technology Management Research 1, 103–118. https://doi.org/10.1016/1047-8310(90)90016-W

Greenberg, J., Mollick, E.R., 2018. Sole Survivors: Solo Ventures Versus Founding Teams, SSRN Electronic Journal. https://doi.org/10.2139/ssrn.3107898

Gulati, R., Higgins, M.C., 2003. Which ties matter when? The contingent effects of interorganizational partnerships on IPO success. Strategic Management Journal 24, 127–144. https://doi.org/10.1002/smj.287

Hadley, B., Gloor, P.A., Woerner, S.L., Zhou, Y., 2018. Analyzing VC Influence on Startup Success: A People-Centric Network Theory Approach, in: Grippa, F., Leitão, J., Gluesing, J., Riopelle, K., Gloor, P. (Eds.), Collaborative Innovation Networks: Building Adaptive and Resilient Organizations. Springer International Publishing, New York, NY, pp. 3–14. https://doi.org/10.1007/978-3-319-74295-3_1

He, W., Guo, L., Shen, J.C., Akula, V., 2016. Social Media-Based Forecasting: A Case Study of Tweets and Stock Prices in the Financial Services Industry. Journal of Organizational and End User Computing. https://doi.org/10.4018/joeuc.2016040105

Hellmann, T., 2000. Venture capitalists: the coaches of Silicon Valley, in: Miller, W., Lee, C.M., Hanock, M.G., Rowen, H. (Eds.), The Silicon Valley Edge: A Habitat for Innovation and Entrepreneurship. Stanford University Press, Stanford, CA, pp. 267–294.

Hellmann, T., Puri, M., 2002. Venture Capital and the Professionalization of Start-Up Firms: Empirical Evidence. The Journal of Finance 57, 169–197. https://doi.org/10.1111/1540-6261.00419

Helmers, C., Rogers, M., 2011. Does patenting help high-tech start-ups? Research Policy 40, 1016–1027. https://doi.org/10.1016/j.respol.2011.05.003

Higashide, H., Birley, S., 2002. The consequences of conflict between the venture capitalist and the entrepreneurial team in the United Kingdom from the perspective of the venture capitalist. Journal of Business Venturing 17, 59–81. https://doi.org/10.1016/S0883-9026(00)00057-4

Hillman, A.J., Shropshire, C., Cannella, A.A., 2007. Organizational predictors of women on corporate boards. Academy of Management Journal 50, 941–952. https://doi.org/10.5465/AMJ.2007.26279222

Hite, J.M., 2005. Evolutionary Processes and Paths of Relationally Embedded Network Ties in Emerging Entrepreneurial Firms. Entrepreneurship Theory and Practice 29, 113–144. https://doi.org/10.1111/j.1540-6520.2005.00072.x

Hochberg, Y. V, Ljungqvist, A., Lu, Y., 2007. Whom you know matters: Venture capital networks and investment performance. Journal of Finance 62, 251–301.

https://doi.org/10.1111/j.1540-6261.2007.01207.x

Jin, F., Wu, A., Hitt, L., 2017. Social Is the New Financial: How Startup Social Media Activity Influences Funding Outcomes. Academy of Management Proceedings 2017, 13329. https://doi.org/10.5465/AMBPP.2017.13329abstract

Kaplan, S.N., Stromberg, P., 2003. Financial Contracting Theory Meets the Real World: An Empirical Analysis of Venture Capital Contracts". Review of Economic Studies. Review of Economic Studies 70, 281–315.

Khanin, D., Turel, O., 2013. Conflicts Between Venture Capitalists and CEO's of their Portfolio Companies. Journal of Small Business Strategy 23, 31–54.

Kibler, E., Fink, M., Lang, R., Muñoz, P., 2015. Place attachment and social legitimacy: Revisiting the sustainable entrepreneurship journey. Journal of Business Venturing Insights. https://doi.org/10.1016/j.jbvi.2015.04.001

Kile, C.O., Phillips, M.E., 2009. Using industry classification codes to sample high-technology firms: Analysis and recommendations. Journal of Accounting, Auditing and Finance 24, 35–58. https://doi.org/10.1177/0148558X0902400104

Klotz, A.C., Hmieleski, K.M., Bradley, B.H., Busenitz, L.W., 2014. New Venture Teams: A Review of the Literature and Roadmap for Future Research The NVT Domain. Journal of Management 40, 226–255. https://doi.org/10.1177/0149206313493325

Kortum, S., Lerner, J., 2000. Assessing the Contribution of Venture Capital to Innovation. The RAND Journal of Economics 31, 674. https://doi.org/10.2307/2696354

Kuhn, K., Galloway, T., Collins-Williams, M., 2016. Near, far, and online: small business owners' advice-seeking from peers. Journal of Small Business and Enterprise Development. https://doi.org/10.1108/JSBED-03-2015-0037

Lahr, H., Mina, A., 2016. Venture capital investments and the technological performance of portfolio firms. Research Policy 45, 303–318.

https://doi.org/10.1016/j.respol.2015.10.001

Lasch, F., Le Roy, F., Yami, S., 2007. Critical growth factors of ICT start-ups. Management Decision 45, 62–75. https://doi.org/10.1108/00251740710718962

Li, X., Xie, Q., Jiang, J., Zhou, Y., Huang, L., 2018. Identifying and monitoring the development trends of emerging technologies using patent analysis and Twitter data mining: The case of perovskite solar cell technology. Technological Forecasting and Social Change. https://doi.org/10.1016/j.techfore.2018.06.004

Lin, B.W., Li, P.C., Chen, J.S., 2006. Social capital, capabilities, and entrepreneurial strategies: A study of Taiwanese high-tech new ventures. Technological Forecasting and Social Change. https://doi.org/10.1016/j.techfore.2004.12.001

Liu, L., Wu, J., Li, P., Li, Q., 2015. A social-media-based approach to predicting stock comovement. Expert Systems with Applications 42, 3893–3901. https://doi.org/10.1016/j.eswa.2014.12.049

Lotti, F., Santarelli, E., 2004. Industry Dynamics and the Distribution of Firm Sizes: A Nonparametric Approach. Southern Economic Journal 70, 443. https://doi.org/10.2307/4135325

Lungeanu, R., Zajac, E.J., 2016. Venture capital ownership as a contingent resource: How owner-firm fit influences ipo outcomes. Academy of Management Journal. https://doi.org/10.5465/amj.2012.0871

Mak, Y.T., Kusnadi, Y., 2005. Size really matters: Further evidence on the negative relationship between board size and firm value. Pacific Basin Finance Journal 13, 301–318. https://doi.org/10.1016/j.pacfin.2004.09.002

Mullin, J., 2014. VC Funding Can Be Bad For Your Start-Up [WWW Document]. Harvard Business Review. URL https://hbr.org/2014/08/vc-funding-can-be-bad-for-your-start-up

Myint, Y.M., Vyakarnam, S., New, M.J., 2005. The effect of social capital in new venture creation: the Cambridge high-technology cluster. Strategic Change 14, 165–177. https://doi.org/10.1002/jsc.718

Phillips, F., 2018. The sad state of entrepreneurship in America: What educators can do about it. Technological Forecasting and Social Change 129, 12–15. https://doi.org/10.1016/j.techfore.2018.01.001

Pirolo, L., Presutti, M., 2010. The Impact of Social Capital on the Start-ups' Performance Growth. Journal of Small Business Management 48, 197–227. https://doi.org/10.1111/j.1540-627X.2010.00292.x

Post, C., Byron, K., 2015. Women on boards and firm financial performance: A meta-analysis. Academy of Management Journal. https://doi.org/10.5465/amj.2013.0319

Raz, O., Gloor, P.A., 2007. Size Really Matters—New Insights for Start-ups' Survival. Management Science 53, 169–177.

Reuer, J.J., Devarakonda, R., 2017. Partner selection in R&D collaborations: Effects of affiliations with venture capitalists. Organization Science. https://doi.org/10.1287/orsc.2017.1124

Sapienza, H.J., 1992. When do venture capitalists add value? Journal of Business Venturing 7, 9–27. https://doi.org/10.1016/0883-9026(92)90032-M

Sapinski, J.P., Carroll, W.K., 2017. Interlocking Directorates and Corporate Networks. Handbook of the International Political Economy of the Corporation 1, 1–23. https://doi.org/10.17605/OSF.IO/7T8C9

Shane, S., Cable, D., 2002. Network Ties, Reputation, and the Financing of New Ventures. Management Science 48, 364–381. https://doi.org/10.1287/mnsc.48.3.364.7731

Sigfusson, T., Chetty, S., 2013. Building international entrepreneurial virtual networks in cyberspace. Journal of World Business. https://doi.org/10.1016/j.jwb.2012.07.011

Smith, C., Smith, J.B., Shaw, E., 2017. Embracing digital networks: Entrepreneurs' social

capital online. Journal of Business Venturing 32, 18–34. https://doi.org/10.1016/j.jbusvent.2016.10.003

Spender, J.-C., Corvello, V., Grimaldi, M., Rippa, P., 2017. Startups and open innovation: a review of the literature. European Journal of Innovation Management. https://doi.org/10.1108/EJIM-12-2015-0131

Stam, W., Arzlanian, S., Elfring, T., 2014. Social capital of entrepreneurs and small firm performance: A meta-analysis of contextual and methodological moderators. Journal of Business Venturing. https://doi.org/10.1016/j.jbusvent.2013.01.002

Stam, W., Elfring, T., 2008. Entrepreneurial orientation and new venture performance: The moderating role of intra- and extraindustry social capital. Academy of Management Journal 51, 97–111. https://doi.org/10.5465/AMJ.2008.30744031

Stanfill, C., Sostheim, J., Cordeiro, N., Klees, D., 2017. PitchBook PE & VC Fundraising Report.

Stuart, T.E., Hoang, H., Hybels, R.C., 1999. Interorganizational Endorsements and the Performance of Entrepreneurial Ventures. Administrative Science Quarterly. https://doi.org/10.2307/2666998

Teten, D., AbdelFattah, A., Bremer, K., Buslig, G., 2013. The Lower-Risk Startup: How Venture Capitalists Increase the Odds of Startup Success. The Journal of Private Equity 16, 7–19. https://doi.org/10.3905/jpe.2013.16.2.007

Venkataraman, S., 1997. The Distinctive Domain of Entrepreneurship Research. Advances in Entrepreneurship, Firm Emergence and Growth 3, 119–138. https://doi.org/10.2139/ssrn.1444184

Wasserman, S., Faust, K., 1994. Social Network Analysis: Methods and Applications. Cambridge University Press, New York, NY. https://doi.org/10.1525/ae.1997.24.1.219

Weill, P., Apel, T., Woerner, S.L., Banner, J.S., 2019. It Pays to Have a Digitally Savvy Board.

MIT Sloan Management Review Spring 201, in press.

Westhead, P., Ucbasaran, D., Wright, M., 2003. Differences Between Private Firms Owned by Novice, Serial and Portfolio Entrepreneurs: Implications for Policy Makers and Practitioners. Regional Studies 37, 187–200. https://doi.org/10.1080/0034340022000057488

Wonglimpiyarat, J., 2016. Exploring strategic venture capital financing with Silicon Valley style. Technological Forecasting and Social Change 102, 80–89. https://doi.org/10.1016/j.techfore.2015.07.007

Zahra, S., 1996. Technology strategy and new venture performance: A study of corporate-sponsored and independent biotechnology ventures. Journal of Business Venturing 11, 289–321.

Zhang, X., Fuehres, H., Gloor, P.A., 2011. Predicting Stock Market Indicators Through Twitter “I hope it is not as bad as I fear.” Procedia - Social and Behavioral Sciences 26, 55–62. https://doi.org/10.1016/j.sbspro.2011.10.562